\documentclass[aps,pra,twocolumn,superscriptaddress,groupedaddress]{revtex4}  % for review and submission

\usepackage[paperwidth=210mm,paperheight=297mm,centering,hmargin=1.5cm,vmargin=3.0cm]{geometry}
\usepackage{graphicx}  % needed for figures
\usepackage{dcolumn}   % needed for some tables
\usepackage{bm}        % for math
\usepackage{amssymb}   % for math
\usepackage[T1]{fontenc}

\begin{document}

% The following information is for internal review, please remove them for submission
\widetext
%\leftline{Version xx as of \today}
%\leftline{Primary authors: Joe E. Physics}
%\leftline{To be submitted to (PRL, PRD-RC, PRD, PLB; choose one.)}
%\leftline{Comment to {\tt d0-run2eb-nnn@fnal.gov} by xxx, yyy}
%\centerline{\em D\O\ INTERNAL DOCUMENT -- NOT FOR PUBLIC DISTRIBUTION}

% the following line is for submission, including submission to the arXiv!!
%\hspace{5.2in} \mbox{Fermilab-Pub-04/xxx-E}

\title{Formation of helical ion chains}
  % D0 authors (remove the first 3 lines
                             % of this file prior to submission, they
                             % contain a time stamp for the authorlist)
                             % (includes institutions and visitors)
\date{\today}

\author{R. Nigmatullin}
\affiliation{Institute of Quantum Physics, Ulm University, Albert-Einstein-Allee
11, D-89069, Germany\\}
\affiliation{Department of Materials, University of Oxford, Oxford OX1 3PH, UK \\}
\author{A. del Campo}
\affiliation{Department of Physics, University of Massachusetts Boston, Boston, MA 02125, USA\\}
\author{G. De Chiara}
\affiliation{Centre for Theoretical Atomic, Molecular and Optical Physics,
School of Mathematics and Physics, Queen's University Belfast, Belfast,
BT7 1NN, UK\\}
\author{G. Morigi}
\affiliation{Theoretische Physik, Universit\"{a}t des Saarlandes, D-66123
Saarbr\"{u}cken, Germany\\}
\author{M. B. Plenio}
\affiliation{Institute of Theoretical Physics, Ulm University, Albert-Einstein-Allee
11, D-89069, Germany\\}
\author{A. Retzker}
\affiliation{Racah Institute of Physics, The Hebrew University of Jerusalem,
Jerusalem, 91904, Givat Ram, Israel}
\begin{abstract}
We study the nonequilibrium dynamics of the linear to zigzag structural
phase transition exhibited by an ion chain confined in a trap with
periodic boundary conditions. The transition is driven by reducing
the transverse confinement at a finite quench rate, which can be accurately
controlled. This results in the formation of zigzag domains oriented
along different transverse planes. The twists between different domains
can be stabilized by the topology of the trap and under laser cooling
the system has a chance to relax to a helical chain with nonzero winding
number. Molecular dynamics simulations are used to obtain a large
sample of possible trajectories for different quench rates. The scaling
of the average winding number with different quench rates is compared
to the prediction of the Kibble-Zurek theory, and a good quantitative
agreement is found. 
\end{abstract}
\maketitle
\section{Introduction} 

Plasmas of singly charged ions can be spatially confined by Paul or Penning traps
\cite{Ghosh1995}. When they are laser cooled to sufficiently low temperatures they self-crystallize into Coulomb crystals \cite{doi:10.1080/00107514.2014.989715}. Coulomb ion crystals have attracted considerable attention as a platform for investigating nonlinear and non-equilibrium dynamics close to criticality. Some examples of the studies of the nonlinear dynamics of ion crystals include the simulation of linear and nonlinear Klein-Gordon fields on a lattice \cite{PhysRevLett.101.260504}, the study of nucleation of topological defects \cite{PhysRevLett.105.075701,Pyka2013,Ulm2013}, dynamics of discrete solitons \cite{1367-2630-15-10-103013,1367-2630-15-9-093003}, dry friction \cite{Benassi2011,1367-2630-13-7-075012,Fogarty2015,Vuletic2014}, as 
as well as proposals to realize models related to energy transport \cite{1367-2630-13-7-075012,Ruiz2014} and synchronization \cite{PhysRevLett.106.143001}. Even though all of the above experiments and proposals are classical, the high degree of isolation of the ion crystals from the surrounding environment implies also the possibility to enter the regime where quantum
mechanical effects must be accounted for to describe critical phenomena \cite{PhysRevLett.101.260504,PhysRevLett.106.010401,PhysRevA.83.032308,refId0}
and where the quantum motion can be utilized for quantum information processing using trapped
ions \cite{Cirac1995,PhysRevLett.104.043004}. This paper focuses on ions but with suitable modifications the ideas can be transported to other systems composed of mutually repelling particles in global confining potentials, for example, ultracold atomic dipoles in quasi two-dimensional potentials \cite{Astrakarchik}. 

In this paper we consider the non-equilibrium statistical mechanics of a chain of
ions, following a quench in the transverse potential frequency that induces a linear to zigzag structural phase transition. The finite rate quench results in the creation of structural defects in the zigzag chain, referred to as kinks in a planar two dimensional system. The scaling of the average number of kinks with the quench rate is predicted by the Kibble-Zurek theory (KZ) \cite{Zurek1985,Campo2014}. The analysis in two dimensions has been previously performed by some of us \cite{PhysRevLett.105.075701,1367-2630-12-11-115003} and subsequently the creation of kinks was studied in several non-equilibrium ion trap experiments \cite{Pyka2013,Ulm2013}, see \cite{Campo2014} for a review. Presently, we consider three dimensional crystals in a trap which is invariant under rotations about the trap axis. In such systems finite rate quench in the transverse potential results in twists in a zigzag, which under periodic boundary conditions can stabilize into helices with non-zero winding numbers. The main objective of the paper is to quantify the scaling relation between the winding number and the quench rate using KZ theory and to verify the prediction using molecular dynamics simulations. We also perform a finite size scaling analysis, extending the previous results on KZ scaling in two dimensional planar crystals \cite{PhysRevLett.105.075701,1367-2630-12-11-115003}.

The paper is organized as follows. Section II introduces the ion crystal system and 
reviews the Ginzburg-Landau theory of the structural linear to zigzag phase transition. In Section III the scaling of defects with quench rate is derived using KZ and finite size scaling theory. Section IV describes the simulation method. In Section V the KZ scaling law obtained using the numerical simulations is compared to the theoretical prediction. Finally the conclusions of the paper are drawn.

\section{Ion crystals and Ginzburg-Landau model}

Kibble-Zurek scaling laws are universal as they depend on the universality class of the phase transition and not on the microscopic dynamics of the system. Thus, in order to derive the scaling laws for the linear to zigzag phase transition, the microscopic theory must first be connected to a coarse-grained Ginzburg-Landau theory. This connection was established analytically in \cite{PhysRevB.77.064111} and this section provides an overview of the theory. 

Charged particles are trapped by either using time varying electric fields (Paul traps) or a combination of electric and magnetic fields (Penning traps). Coulomb crystals in such traps are regular periodic solutions to the equations of motion. A common modelling approximation used in the study of Coulomb crystals is the ponderomotive or pseudopotential theory (PPT), which replaces the time-varying trap potential with a time-independent harmonic potential. For ions in a Paul trap PPT captures the secular motion of the ions but neglect the rapid micromotion. In the current paper, we will always use PPT, since it facilitates the derivation of Ginzburg-Landau theory and the numerical simulations. PPT is a good approximation for the purpose of studying the linear to zigzag phase transition, since it correctly predicts the equilibrium positions of the ions and the vibrational spectrum in the linear chain configuration \cite{1367-2630-14-9-093023,PhysRevLett.109.263003}. The Ginzburg-Landau theory for the linear to zigzag phase transition, which is crucial for the subsequent analysis, relies only on the normal modes and frequencies of the linear chain in the vicinity of the critical point of the structural phase transition.

Assuming PPT, the potential energy of the system consisting
of $N\rightarrow\infty$ ions is given by

\begin{equation}
V=\frac{1}{2}m\omega_{r}^2\sum_{j=1}^{N}\left(x_{j}^2+y_{j}^2\right)+Q^{2}\sum_{i<j}^{N}\frac{1}{|\textbf{r}_{i}-\textbf{r}_{j}|},\label{eq:a}
\end{equation}
where $\textbf{r}_{j}=(x_{j},y_{j},z_{j})$ are the coordinates of
the $j$th ion, $Q^{2}\equiv e^{2}/4\pi\epsilon_{0}$, $e$ is the charge of the ions, $\epsilon_0$  is the vacuum permittivity, $\omega_{r}$
is the radial secular frequency, $m$ is the mass of the ions, 
 $x$ and $y$ denote radial directions and $z$ denotes the axial direction. The potential (\ref{eq:a}) is for a system in thermodynamic limit, where the number of charges is infinite. Above a certain critical value of $\omega_r$ the lowest energy configuration is a single row of evenly spaced particles along the $z$-axis. In the thermodynamic limit the system is translationally invariant and thus homogeneous - the inter-ion spacing is a constant $a$. In most ion trap experiments the axial confinement is achieved using a weak harmonic potential in the $z$-direction and the chain is finite. Crystals in such harmonic traps are inhomogeneous with $a$ varying along the chain.   
Here, we will be dealing solely with the homogeneous crystals i.e. crystals in thermodynamic limit or with periodic boundary conditions. Such homogeneous systems could be realized in the laboratory in ring traps \cite{Birkl1992,PhysRevLett.68.2007} and octupole traps \cite{PhysRevA.75.033409,PhysRevA.81.043410,PhysRevA.87.013425}. 

Ginzburg-Landau (GL) potential is derived by Taylor expanding the potential (\ref{eq:a}) in small displacement around the equilibrium positions of the ions in the linear chain configuration and keeping up to fourth order terms in the radial displacements \cite{PhysRevB.77.064111,1367-2630-12-11-115003}. The result is given by

\begin{equation}
V=\frac{1}{2}\frac{m}{a}\int dz\left[\delta|\phi|^{2}+h^{2}\partial_{z}\phi\partial_{z}\phi^{*}+g|\phi|^{4}\right],\label{eq:b}
\end{equation}
where  $\phi((k+1)a)-\phi(ka)\ll a$, $\delta$,
$h$, $g$ are constants that depend on the trap parameters and $\phi$
is the complex order parameter that is related to the original degrees
of freedom by 

\begin{equation}
\phi(ka)=(-1)^{k}\left(x_{k}+iy_{k}\right).\label{eq:b2}
\end{equation}
where $x_{k}$ and $y_{k}$ are the radial coordinates of the $k$th
ion and $i=\sqrt{-1}$. The field $\phi$ is the transverse
displacement of the ions from the axis, but with the reversed sign
for every odd ion. 

The parameters in GL energy are $h=\omega_0 a \sqrt{\textrm{log}2}$, $g=(93\zeta(5)/32)\omega_0^2/a^2$ and the control parameter, $\delta$, is given by

\begin{equation}
\delta=\omega_{r}^{2}-\left(\omega_{r}^{(c)}\right)^2,\label{eq:b3}
\end{equation}
where $\omega_{r}^{(c)}=\sqrt{7\zeta(3)/2}\omega_{0}$ and $\omega_{0}=\sqrt{e^{2}/4\pi\epsilon_{0}ma^{3}}$.

The GL potential (\ref{eq:b}) arises naturally in physical systems where the order parameter has rotational symmetry. For example near the critical point, symmetry breaking in Josephson tunnel junctions \cite{PhysRevLett.96.180604,PhysRevLett.89.080603} as well as Bose-Einstein condensates \cite{Pitaevskii2003} (within a Gross-Pitaevskii description) have the same dimensionality and symmetry properties as the linear to zigzag transition in ion traps. As a matter of fact, all these systems belong to the mean-field theory universality class.

\begin{figure*}
\centering

\includegraphics[scale=0.9]{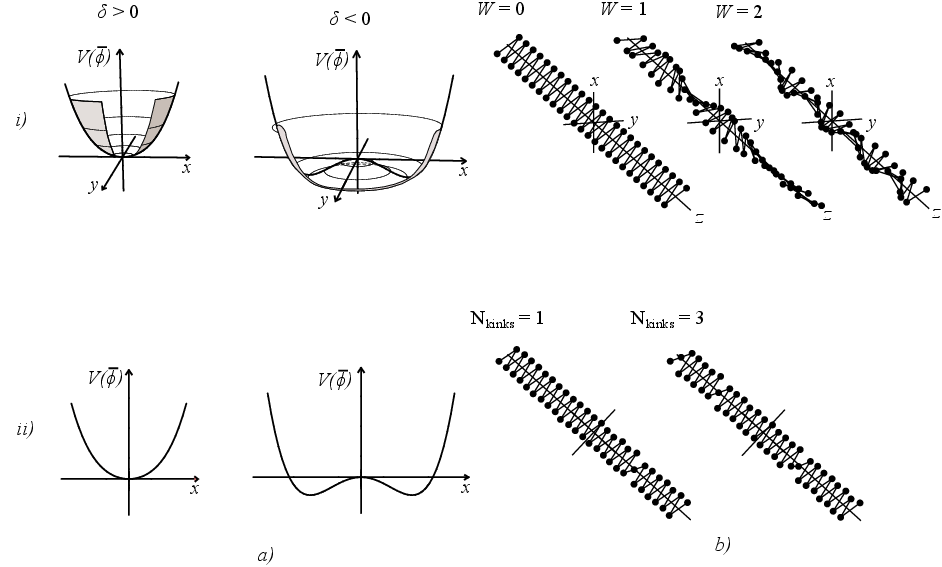}

\caption{a) The potential energy of the mean field configuration $\bar{\phi}$ of the ion chain in the i) three dimensional system and symmetric radial potential and ii) in two the two dimensional system. The potential are shown for cases $\delta>0$ and $\delta<0$; the transition from single well to a double well or Mexican hat potential occurs at $\delta=0$. b) Examples of several stable zigzag chain configurations produced as a result of a non-isothermal quench from a linear to zigzag phase in i) three dimensional system and ii) two dimensional system. Helical configurations of winding numbers 0, 1 and 2 are shown and two dimensional zigzag configurations with 1 and 3 kinks are shown. All of the configurations were found by quenching ion chains across the structural transition using molecular dynamics simulations.}
\end{figure*}

The phase transition exhibited by model (\ref{eq:b}) is a second order symmetry breaking phase transition. When $\delta>0$ the system is in the symmetric state where the lowest energy configuration is $\bar{\phi}=0$
i.e. a linear chain is stable. When $\delta<0$ the system is in the symmetry broken state where the lowest energy configuration is $\bar{\phi}=e^{i\theta}\sqrt{-\delta/2g}$ with
$\theta\in[0,2\pi]$. This corresponds to a zigzag configuration at
an angle $\theta$ to a chosen reference plane. The critical point
of the phase transition is $\delta=0$ or $\omega_{r}=\omega_{r}^{(c)}$.
Figure 1a) illustrates the functional form of the ground state potential energy, $V(\bar{\phi})$, for $\delta>0$ (single well potential) and for $\delta<0$ (Mexican hat potential). Model (\ref{eq:b}) allows for phase winding solutions - these are stable configurations where the phase $\theta$ varies along the crystal. With periodic boundary conditions the total phase must be equal to $2\pi W$, where $W$ is an integer known as the winding number

\begin{equation}
W=\frac{1}{2\pi}\int\partial_{z}\theta(z)dz.\label{eq:W}
\end{equation}

Examples of possible stable configurations with winding numbers of 0 (ground state), 1 and 2 are shown in figure 1. In this paper, configurations of non-zero winding number are referred to as helical structures. One should emphasize that these twisted zigzag structures are different from helical crystal phases that were predicted numerically \cite{Hasse1990419} and observed experimentally \cite{PhysRevLett.68.2007}. One enters the helical structural phase by reducing the radial confining frequency below a critical value at which the zigzag crystal becomes unstable. Topological helices that are subject of the present paper have not yet been observed in experiments.

The strength of confinement in the $y$ and $x$ directions may in general be unequal, in which case the trap is radially asymmetric. In strongly asymmetric traps all of the ions in the crystal lie in one plane. For two dimensional planar crystals equation (\ref{eq:b}) holds, except now the order parameter $\phi$ is real, since all coordinates in one of the transverse directions are zero. In two dimensions, for $\delta>0$ the ground state is $\bar{\phi}=0$, the system is in a linear phase and the potential is a single well (figure 1 b)). For $\delta<0$
the ground state of the system is $\phi=e^{i\bar{\theta}}\sqrt{-\delta/2g}$ with $\theta=0$ or $\theta=\pi$, the ground state is a zigzag chain and the potential is of the form of a double well (figure 1b)). In the symmetry broken phase ($\delta < 0$), the system supports stable kink solutions - solutions where the field interpolates between the two possible ground state values $+\sqrt{-\delta/2g}$ and $-\sqrt{-\delta/2g}$. The number of kinks in the system is defined as the number of times $\phi$ crosses the $z$-axis. 
Figure 1b)ii) shows examples of one and three kinks in chains where particles experience full Coulomb interactions. This type of structural defects is often referred to as $Z_2$ kinks or solitons, since they arise as a result of phase transitions that break reflectional $Z_2$ symmetry. Kinks in two dimensional Coulomb crystals were studied theoretically and experimentally as discrete soliton model systems \cite{PhysRevLett.68.2007,1367-2630-15-9-093003,PhysRevLett.110.133004}, as possible qubit candidates for quantum information processing \cite{PhysRevLett.104.043004} and in the context of KZ mechanism \cite{PhysRevLett.105.075701,1367-2630-12-11-115003,Pyka2013,Ulm2013}.

\section{Non-equilibrium dynamics and scaling laws}

Near the critical point, the quench dynamics can be modelled by the time dependent Ginzburg-Landau equation \cite{1367-2630-12-11-115003}

\begin{equation}
\partial^2_{tt}\phi+\Gamma\partial_t\phi+\delta\phi+h^{2}\partial^2_{zz}\phi+g|\phi|^{2}\phi=\theta(z,t),\label{eq:d}
\end{equation}
where $\partial^2_{tt}\phi$ is the inertial term, $\Gamma\partial_t\phi$ is the friction term and $\theta(z,t)$
is the stochastic terms. Equation
of motion for the complex conjugate field $\phi^{*}$ is analogous
to equation (\ref{eq:d}). The stochastic force satisfies the following statistical relationships

\begin{eqnarray}
\left\langle \theta(z,t)\right\rangle  & = & 0\label{eq:e}\\
\left\langle \theta_{\alpha}(z,t)\theta_{\beta}(z',t')\right\rangle  & = & 2\Gamma k_{B}T\delta_{\alpha\beta}\delta(z-z')\delta(t-t').\label{eq:f}
\end{eqnarray}
where $\left\langle...\right\rangle$ denotes the ensemble average.
The Langevin dynamics given by (\ref{eq:d})-(\ref{eq:f})
simulates the system in contact with the thermal bath at temperature $T.$ In ion traps the friction and stochastic terms arise because of the interactions between the ions and the Doppler cooling laser beam.

Suppose that the radial frequency $\omega_r$ is externally varied such as to induce a linear quench in $\delta$ at a rate proportional to $v$

\begin{equation}
\delta(t)=-\delta_0 v t \; \textrm{sign}(t),\label{eq:g}
\end{equation}
where $t\in [-t_0,t_f]$, $t_0>0$ and $t_1>0$. The quench rate $v$ is made dimensionless by taking $v=1/(\tau_Q \omega_0)$, where $\tau_Q$ is the quench time and  $ \omega_0=\sqrt{e^2/4 \pi \epsilon_0 m a^3}$. The value of $t_0$ is taken to be large enough so that the system is far from the critical point and the correlation length equals to the microscopic length scale $\xi\sim \delta^{-\nu}\rightarrow a$ i.e. there should be no long range correlations. 

The finite rate quench drives the system out of equilibrium and as a result there is a finite probability that at the end of the quench the system will contain a number of stable defects. Qualitatively, it is expected that the slower the quench the less defects will form. In the limit of infinitely slow quenches the dynamics is isothermal and the final state always belongs to the lowest energy ground state manifold. At finite quench rates, a system can undergo phase transition faster than the time it takes for phonons (information) to propagate across the whole system. Causally disconnected regions select the ground state independently and this lack of coordination results in topological defects. Faster quenches result in more causally disconnected regions and hence more topological defects.

The quantitative scaling law relating the number of defects and the quench rate is established by KZ theory \cite{Kibble2007,Zurek1985,Zurek1993}. KZ theory connects the important length and time scale in the system during the non-equilibrium quench to the characteristic quench time $1/v$. This time and length scale are often referred to as ``freeze-out time'' $\hat{t}_L$ and ``freeze-out correlation length'' $\hat{\xi}_L$, where $L$ refers to the size of the system. The spatial characteristics of the system such as the density of defects are then related to the ``freeze-out correlation length'' and hence $v$. In the thermodynamic limit of infinite systems, both $\hat{\xi}_\infty$ and $\hat{t}_\infty$ entirely depend on the critical exponents associated with the symmetry breaking phase transition. According to KZ theory, $\hat{\xi}_\infty$ is given by 

\begin{equation}
\hat{\xi}_{\infty}\sim v^{-\frac{\nu}{1+\nu z}},
\label{eq:KZscaling}
\end{equation}
where $\nu$ is the critical exponent associated with the divergence of correlation length at equilibrium, $\xi\sim\delta^{-\nu}$, and $z$ is the critical exponent associated with the divergence of relaxation time observed for small perturbation from equilibrium, $\tau\sim\delta^{-z \nu}$. For the mean field universality class $\nu=1/2$, and for underdamped dynamical regime, which is relevant for ion crystal dynamics, the dynamic critical exponent is $z=1$ and hence $\hat{\xi}_\infty \sim v^{-1/3}$ \cite{1367-2630-12-11-115003,PhysRevLett.78.2519,PhysRevD.58.085021}. We may also obtain KZ scaling by directly rescaling the length and time such that the equations of motion become independent of $v$, thereby identifying the natural length and time scale in the dynamics \cite{Nikoghosyan}. If we neglect the non-linear interaction term, which is small near the critical point, and the forces due to laser cooling (for the underdamped dynamics) the equation of motion reads

\begin{equation}
\frac{\partial^{2}\phi}{\partial t^{2}}-\delta_0 v t\;\textrm{sign}(t) \phi+h^{2}\frac{\partial^{2}\phi}{\partial z^{2}}=0,\label{eq:h}
\end{equation}
where the system is taken to be infinite i.e. $z\in(-\infty,\infty)$.

Consider a linear rescaling of $z$ and $t$

\begin{eqnarray}
Z & = & z/\hat{\xi}_\infty\label{eq:g1}\\
T & = & t/\hat{t}_\infty,\label{eq:g3}
\end{eqnarray}
where $\hat{\xi}_\infty$ and $\hat{t}_\infty$ are the sought scaling factors. 
Substituting (\ref{eq:g1}) and (\ref{eq:g3}) in (\ref{eq:h}) gives 

\begin{equation}
\frac{\partial^{2}\phi}{\partial T^{2}}-\eta\delta_{0} \hat{t}^3 \textrm{sign}(T) \phi+h^{2} \frac{\hat{t}^2}{\hat{\xi}^2}\frac{\partial^{2}\phi}{\partial Z^{2}}=0.\label{eq:g5-1}
\end{equation}

The equation (\ref{eq:h}) becomes independent of $v$ with the following choice of $\hat{\xi}$ and $\hat{t}$

\begin{eqnarray}
\hat{\xi}_\infty & \sim &v^{-1/3}\label{eq:h1}\\
\hat{t}_\infty & \sim & v^{-1/3},\label{eq:h2}
\end{eqnarray}
The rescaling of the spatial and temporal variables in the quench
equation (\ref{eq:h}) according to (\ref{eq:g1})-(\ref{eq:g3})
brings the equation into $v$-independent and hence universal
form. The important length and time scale during the quench are, therefore, $\hat{\xi}_\infty\sim v^{-1/3}$ and $\hat{t}_\infty\sim v^{-1/3}$ i.e. the same scaling relations that arise from equation (\ref{eq:KZscaling}) with $\nu=1/2$ and $z=1$. 

Finite systems with periodic boundary conditions accurately model infinite systems in thermodynamic limit as long as the correlation length is significantly smaller than the size of the system. Thus we expect that $\hat{\xi}_L$ in finite systems is approximately equal to $\hat{\xi}_\infty$ for some range of quench rates, but at very slow quenches the correlation length may become large enough to be comparable to the system size $L$ and the boundary effects would modify the KZ prediction.  We now develop the finite size KZ scaling theory for linear to zigzag structural transition following the treatment presented in \cite{Kolodrubetz2012} for non-equilibrium quantum transition from paramagnetic to antiferromagnetic phase. The crossover from the KZ scaling in thermodynamic limit to a finite-size scaling is expected to occur when $\hat{\xi}_\infty\sim L$. Using the equation (\ref{eq:KZscaling}) and taking the system size $L$ to be proportional to the number of ions $N$, the crossover to finite-size scaling occurs at a critical quench rate $v^{(c)}\sim N^{-(1/\nu+z)}$. Accordingly, we postulate a finite size scaling relation $\hat{\xi}_N\sim \hat{\xi}_{\infty}f\left(v/v^{(c)}\right)$ i.e.

\begin{equation}
\hat{\xi}_N\sim v^{-\frac{\nu}{1+\nu z}} f\left(N^{\frac{1}{\nu}+z} v\right),
\end{equation}
where the scaling function $f(x)$ has the following asymptotic behaviour, $f(x\gg1)\sim \text{constant}$ (KZ regime) and $f(x\ll 1)\sim x^{\frac{\nu}{1+\nu z}}$ (slow nearly isothermal quenches that do not generate defects). In practice the scaling should break down at very fast quenches, where the correlation length is comparable to the microscopic length scale (inter-ion spacing), and in which case the GL hydrodynamic theory is no longer valid.

We now address the question of how the number of topological defects depend on $\hat{\xi}_L$ and to determine the scaling with $v$.  In the case of the two dimensional system, the average distance between kinks is simply proportional to $\hat{\xi}_L$ and the expected number of domains is $\left\langle N_d\right\rangle \sim L/\hat{\xi}_L$ i.e.

\begin{equation}
\left\langle N_d\right\rangle\sim N v^{\frac{\nu}{1+\nu z}} g\left(N^{\frac{1}{\nu}+z}v\right), \label{eq:nd_scaling}
\end{equation}
where the function $g$ is the reciprocal of $f$.

In the three dimensional case and the helix formation, the relation between the winding number distribution and quench rate can be obtained by an argument which was used to derive the KZ scaling of the winding number of Bose-Einstein-Condensate wavefunction obtained by a nonequilibrium quench in a radially symmetric toroidal trap \cite{Das2012}. This argument is also used to predict the winding numbers in strongly coupled holographic superconductors described by gauge-gravity duality \cite{Sonner2015}. One assumes that the chain of length
$L$ is divided into $n=L/\hat{\xi}$ regions and each region picks
at random an orientation $\theta$ to some fixed reference plane.
Thus there are $n$ random variables, each uniformly distributed between
$0$ and $2\pi$ and thus having a mean of zero and a variance of
$\pi^{2}/3$. The winding number is $W\approx\frac{1}{2\pi}\sum\theta_{j}$
and hence the distribution of the winding number is a convolution
of $n$ uniformly distributed random variables. For large $n$ the
central limit theorem is valid and $W$ will have a Gaussian distribution
with mean zero and variance $L\pi^{2}/(2\hat{\xi})$. Thus $\left\langle W^{2}\right\rangle$ scales in the same way as $\left\langle N_d\right\rangle$ i.e  $\left\langle W^{2}\right\rangle\sim N/\hat{\xi}_L$. Substituting  $\nu=1/2$ and $z=1$ into equation (\ref{eq:nd_scaling}) gives the following scaling relations for the observables
\begin{eqnarray}
\left\langle N_d\right\rangle &\sim& N v^{\frac{1}{3}} g\left(N^3 v\right),\\ \label{eq:Nd}
\left\langle W^{2}\right\rangle &\sim&N v^{\frac{1}{3}} g\left(N^3 v\right). \label{eq:Wsq}
\end{eqnarray}

\section{Simulation method}
We use molecular dynamics simulations to obtain a large number of trajectories ($\sim$2000) of ion crystals undergoing quenches at different rates from a linear to zigzag configurations. This allows us to determine the average number of defects for a given quench rate and hence the KZ scaling. In KZ studies the simulations often involve numerically solving the hydrodynamic equations such as Gross-Pitaevskii
equation \cite{PhysRevLett.104.160404,PhysRevLett.106.135301} or the time dependent Ginzburg-Landau equation \cite{PhysRevLett.78.2519,PhysRevD.58.085021}.
In contrast here, we simulate the underlying microscopic equations. The
coarse-grained field description of the system, given in the previous
section, is used only to derive the expected scaling. 

The equations of motion for the $j$th ion are given by

\begin{eqnarray}
m\partial_{tt}x_{j} & = & -m\omega(t)^{2}x_{j}-\Gamma\partial_{t}x_{j}-\partial_{x_{j}}V_{c}+\theta_{xj}(t),\label{eq:sim1}\\
m\partial_{tt}y_{j} & = & -m\omega(t)y_{j}-\Gamma\partial_{t}y_{j}-\partial_{y_{j}}V_{c}+\theta_{yj}(t),\\
m\partial_{tt}z_{j} & = & -\partial_{z_{j}}V_{c}+\theta_{zj}(t),
\end{eqnarray}
where $m$ is the mass of the ion, $\omega(t)$ is the transverse
confining frequency, $V_{c}$ is the Coulomb potential energy, $\Gamma$
is the friction coefficient and $(\theta_{xj},\theta_{yj},\theta_{zt})$
is the stochastic thermal force acting on the $j$th ion. The simulated
system is periodic - the axial coordinates of all of the ions are
restricted to the region of $[-L/2,L/2]$. 

The quench is chosen to be such that the transverse frequency is decreased linearly from an initial value $\omega_i$ to the final value $\omega_f$. The time $\tau_Q$ taken for the transverse frequency to reach the final value is varied from experiment to experiment. In the KZ experiments slow quenches are used and the variation in $\delta$ (equation (\ref{eq:g})) is approximately linear. At the start of each quench the system is evolved at constant trap parameters for 200 $\mu$s in order to initialize the system in thermal equilibrium.

All of the simulation were carried out using Langevin-Impulse integration
method \cite{Skeel2002}. The following parameters were used. The mass of
the ions was set to $m=172$ amu, which corresponds to Yb$^{+}$ ions.
The spacing between ions in the linear chains was set to $a=12.9$ $\mu$m
giving $\omega_0=\sqrt{e^2/4 \pi \epsilon_0 m a^3}=610$ kHz.
 Temperature was set to $T=5$ mK and friction coefficient
to $\Gamma=1.5\times10^{-21}$ \mbox{kg s$^{-1}$} obtained by assuming optimal
Doppler cooling on the $^{2}S_{1/2}-$$^{2}P_{1/2}$ transition. The
secular frequencies of the confining potential in the $y$ direction
were $\omega_{xi}/(2\pi)=239$ kHz and $\omega_{xf}=140$ kHz. In
the case of the two dimensional experiment the confining potential
in the $y$ direction was set to a constant value of $\omega_{y}/(2\pi)=477$
kHz. In the case of the three dimensional experiment the confining
potential in the $y$ direction was set to be equal to the confining
potential in the $x$ direction at all times. The quench times ranged
from around around 60 $\mu$s to 2 ms. The integration timestep was
set to 3.2 ps.

\section{Simulation results and discussion}

\begin{figure*}
\centering

\includegraphics{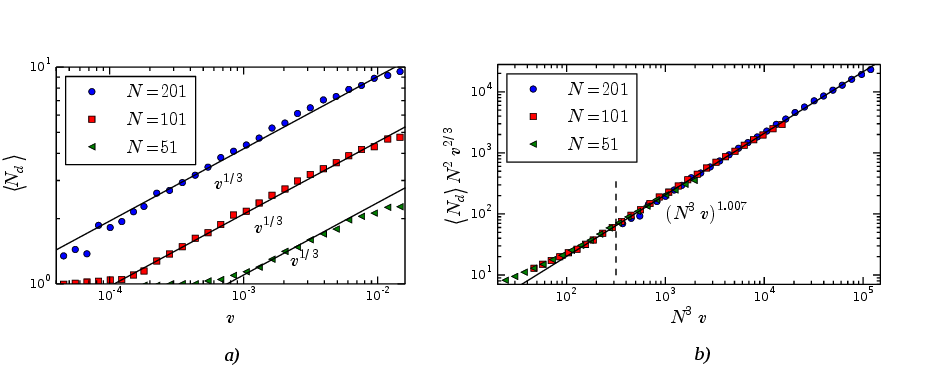}

\caption{a) Scaling of the average number of domains $\langle N_d \rangle$ as a function of quench rates for two dimensional crystals consisting of 51, 101 and 201 ions. The lines indicate the theoretically predicted 1/3 scaling law. b) The plot of $\langle N_d \rangle N^2 v^{2/3}$ versus $N^3 v$ and the collapse of the three curves.  The black solid line was obtained by performing a linear regression fit of the combined data in the range  $N^3 v > 316.0$ ($N^3 v = 316.0$ is indicated by a dashed line).
\label{fig:2dKZscaling}}
\end{figure*}

\begin{figure*}

\includegraphics{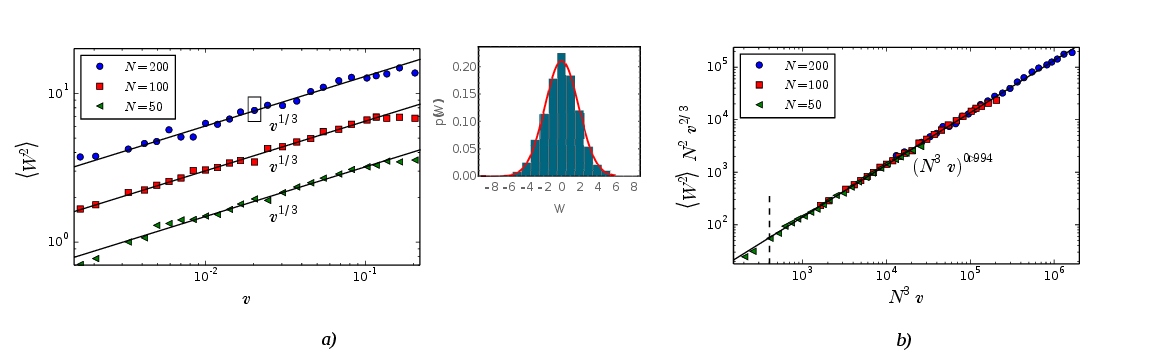}

\caption{Scaling of the variance of the winding number distribution $\left\langle W^{2}\right\rangle$ as a function of quench rate evaluated for three dimensional chains of 200, 100 and 50. The lines indicate the theoretically predicted 1/3 scaling law.  The inset displays the measured winding number distribution for a data set highlighted by a rectangular box on the graph. The red curve in the inset is a Gaussian distribution with mean zero and the variance of the winding number distribution. b) The plot of $\langle W^2 \rangle N^2 v^{2/3}$ versus $N^3 v$ and the collapse of the three curves.  The black solid line was obtained by performing a linear regression fit of the combined data in the range  $N^3 v > 403.0$ ($N^3 v = 403.0$ is indicated by a dashed line). \label{fig:3dKZscaling}}
\end{figure*}

\begin{figure}
\centering

\includegraphics[scale=0.5]{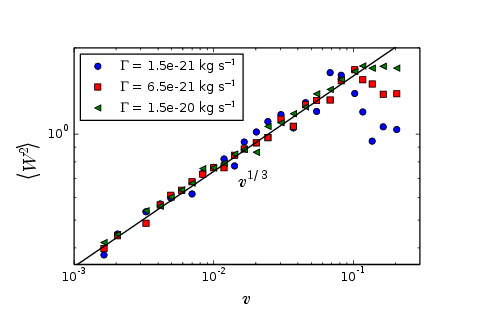}

\caption{Scaling of the variance of the winding number distribution $\left\langle W^{2}\right\rangle$ obtain for chains of 100 ions using simulations with three different friction coefficients $\Gamma = 1.5\times 10^{-20}$ kg s$^{-1}$,  $\Gamma = 1.5\times 10^{-21}$ kg s$^{-1}$ and $\Gamma = 6.5\times 10^{-20}$ kg s$^{-1}$. The solid line indicates a theoretically predicted 1/3 scaling law.  \label{fig:3dKZscalingFriction}}
\end{figure}
Figure \ref{fig:2dKZscaling} shows the results of two dimensional simulations that were done for ion chains containing
51, 101 and 201 ions. In all of the experiments the number of domains $N_d$ is counted at the end of the simulation. Figure \ref{fig:2dKZscaling}a) shows the average number of domains $\langle N_d \rangle$  as a function of quench rate $v$ on a logarithmic scale. In the graph one observes a strong indication of the expected scaling behaviour of $\langle N_d \rangle \sim v^{1/3}$, a plateau at slow quenches due to the finite size effect and a plateau at fast quench rates. To verify the scaling more precisely we plot $\langle N_d \rangle N^2 v^{2/3}$ as a function of $N^3 v$. Equation (\ref{eq:Nd}) suggests that in such a plot 
the three curves collapse onto a universal quench function $xg(x)$, which is indeed clearly visible in figure 2b) even at slow quenches where there are deviation from the thermodynamic limit KZ scaling law. To the best of our knowledge this is a first demonstration of finite size KZ scaling in a classical non-equilibrium thermodynamic process. A comparison of the measured scaling to the predicted scaling of 1/3 is obtained by making linear regression fit in the range $N^3 v > 316.0$, which gives $\langle N_d \rangle N^2 v^{2/3} \propto \left( N^3 v \right)^{1.007}$ and hence $\langle N_d \rangle \propto v^{0.3403}$. The deviation of the measured exponent from the theoretical prediction of 1/3 is 2$\%$.

Figure \ref{fig:3dKZscaling} displays the results of the three dimensional
experiments for the chains of 50, 100 and 200 ions. The winding number of the helices $W$ is determined in the end of each simulation. A plot of $\langle W^2 \rangle$ versus $v$ is shown in figure \ref{fig:3dKZscaling}a). An inset contains a histogram of a selected distribution of $W$ and a probability density function of a normal distribution with the same variance. A close match between the histogram and the Gaussian justifies the use of the central limit theorem in section III for the derivation of the scaling law for $\langle W^2 \rangle$.  
The results shown in figure \ref{fig:3dKZscaling}a) are in good agreement with the predicted scaling of $\langle W^2 \rangle \propto v^{1/3}$. For an accurate quantification of the scaling behaviour we plot $\langle W^2 \rangle N^2 v^{2/3}$ versus $N^3 v$ to collapse the three curve as suggested by equation (\ref{eq:Wsq}).  This plot is shown in figure  \ref{fig:3dKZscaling}b) where indeed the collapse of the curves is clearly visible. To obtain the scaling exponents that approximates the exponent in thermodynamic limit we perform a linear regression fit in the region $N^3 v > 403.0$, that avoids finite size effects at slow quenches. The fitted scaling is $\langle W^2 \rangle N^2 v^{2/3} \propto \left( N^3 v \right)^{0.994}$, which implies $\langle W^2 \rangle \propto v^{0.3273}$. The measured scaling exponent deviates from the predicted exponent of 1/3 by 1.8 $\%$. More data at slow quenches is needed in order to quantify precisely the modifications of the scaling law by the finite size effect. We leave the investigation of the finite size effects in non-equilibrium $U(1)$ symmetry breaking processes for future investigations.

In the underdamped dynamical regime the average number of defects should not depend on the friction coefficient $\Gamma$ \cite{PhysRevD.58.085021}. However, if we increase $\Gamma$, at certain point the frictional force will start to dominate and the dynamics will be overdamped with a different KZ scaling. In order to verify that the system is indeed underdamped and the scaling law is not sensitive to $\Gamma$, quenches at three different friction coefficients are simulated in a system composed of 100 ions. The three friction coefficients used in the simulations are  $\Gamma = 1.5\times10^{-20}$ kg s$^{-1}$, $\Gamma = 6.5\times10^{-21}$ kg s$^{-1}$ and $\Gamma = 1.5\times10^{-21}$ kg s$^{-1}$. Figure \ref{fig:3dKZscalingFriction} shows that in the KZ scaling regime there is no statistically significant difference between the results of the simulations with these three different friction coefficients, which confirms the validity of the underdamped model. Interestingly, we can see that at fast quench rates, where one typically expects to see a plateau, there is a consistent decrease of the number of helices with increasing quench rate. This antiKZM behaviour is more pronounced at small friction coefficients. A possible reason for this is that if there is a large amount of undissipated kinetic energy in the system, the topological defects are very mobile and frequently annihilate one another. An experimental observation of antiKZ scaling in a  system driven through a ferroelectric phase transition was reported in \cite{Griffin2002}. It is possible that there is a common origin between the antiKZM behaviour seen in figure \ref{fig:3dKZscalingFriction} and in  \cite{Griffin2002}. We leave a systematic investigation of antiKZM as a subject for future work.

\section{Conclusions} 

In this paper we examined the non-equilibrium dynamics of Coulomb crystals undergoing a structural transition from a linear to zigzag configuration in rotationally symmetric homogeneous traps. It was shown using Ginzburg-Landau theory that this is a $U(1)$ symmetry breaking phase transition. The symmetry broken zigzag phase supports stable phase winding solutions, which are referred to as helical structures. The probability of obtaining a helix of a certain winding number depends on the quench rate of the transition. We have applied the universal Kibble-Zurek theory to derive the scaling law connecting the variance of the winding number distribution and the quench rate in the underdamped dynamical regime. The scaling law was verified using extensive molecular dynamics simulations of quenches in chains of three different sizes. 
A good quantitative agreement between the results of the simulations and the predictions of the Kibble-Zurek theory was found using finite size theory analysis. The scaling was shown to be robust to the variations of system size and friction coefficient. At fast quench rates and small friction coefficient,
we observed an intriguing antiKZM behaviour in the scaling of the winding number.

This work shows once more that ion crystals are
very well suited classical simulators of complex and critical dynamics. We hope that this work will stimulate progress towards the experimental observations of the predicted helical crystals and further investigations of non-equilibrium phenomena in Coulomb crystals.

\emph{Acknowledgments. }It is a pleasure to thank Bogdam Damski, Arnab Das, and Wojciech H. Zurek for
fruitful comments and discussions. We gratefully thank the bwGRiD project \cite{bwGRID} for the computational resources. RN was supported by EPSRC National Quantum Technology Hub in
Networked Quantum Information Processing and by grant from
the Ministry of Science, Research and the Arts of Baden-W\"{u}rttemberg
(Az: 33-7533-30-10/19/2). AD is supported by the U.S. Department of
Energy through the LANL/LDRD Program. This work was supported by DFG SFB TRR/21, the EU Integrating project SIQS, the EU STREP project EQUAM and an Alexander von Humboldt Professorship.

\appendix

\end{document}